\documentclass[aps,pra,bibnotes,twocolumn]{revtex4}
\usepackage{graphicx}
\usepackage{amsmath,amssymb}

\newcommand{\Sr}{\ensuremath{^{88}}Sr}

\begin{document}

\title{Optical Feshbach resonances: Field-dressed Theory and comparison with experiments}
\author{T.L. Nicholson}
\altaffiliation{Present Address: Center for Ultracold Atoms, Massachusetts Institute of Technology, Cambridge, MA 02139, USA}
\author{S. Blatt}
\altaffiliation{Present Address: Department of Physics, Harvard University, Cambridge, MA 02138, USA}
\author{B.J. Bloom}
\altaffiliation{Present Address: Intel, Hillsboro, OR, USA}
\author{J.R. Williams}
\altaffiliation{Present Address: Jet Propulsion Laboratory, California Institute of Technology, Pasadena, CA 91109, USA}
\author{J.W. Thomsen}
\altaffiliation{Permanent Address: The Niels Bohr Institute, Universitetsparken 5, 2100 Copenhagen, Denmark}
\author{J. Ye}
\affiliation{JILA, National Institute of Standards and Technology and University of Colorado,
Department of Physics, University of Colorado, Boulder, Colorado 80309-0440, USA}
\author{Paul S. Julienne}
\affiliation{Joint Quantum Institute, University of Maryland and NIST, Room 2103 CSS Building, College Park, Maryland 20742, USA}

\date{\today}

\begin{abstract}
Optical Feshbach resonances (OFRs) have generated significant experimental interest in recent years. These resonances are promising for many-body physics experiments, yet the practical application of OFRs has been limited. The theory of OFRs has been based on an approximate model that fails in important detuning regimes, and the incomplete theoretical understanding of this effect has hindered OFR experiments. We present the most complete theoretical treatment of OFRs to date, demonstrating important characteristics that must be considered in OFR experiments and comparing OFRs to the well-studied case of magnetic Feshbach resonances. We also present a comprehensive treatment of the approximate OFR model, including a study of the range of validity for this model. Finally, we derive experimentally useful expressions that can be applied to real experimental data to extract important information about the resonance structure of colliding atoms.
\end{abstract}

\maketitle

\section{Introduction}
\subsection{Background}
\label{sec:intro}

Magnetic Feshbach resonances (MFRs) have become a staple of quantum gas experiments with alkali-metal atoms, allowing for unprecedented control of interatomic interactions \cite{Chin2010}. The MFR technique is so powerful that it has extended the reach of dilute quantum gas experiments to a variety of areas of physics. Examples of high impact experiments that utilize MFRs are the study of strongly correlated systems \cite{Bloch2008} such as unitary Bose \cite{Rem2013,Wild2012} and Fermi \cite{Kinast2004,Regal2004,Zwierlein2004} gases, the discovery of exotic few-body bound states \cite{Kraemer2006,Braaten2006,Ferlaino2011,Wang2013}, the ability to make ultracold molecules~\cite{Ni2008,Quemener2012}, and the engineering of novel quantum matter \cite{Chotia2012,Yan2013}.  Feshbach resonances based on laser fields---known as ``optical Feshbach resonances'' (OFRs) \cite{Fedichev1996,Bohn1997}---have also been observed \cite{Fatemi2000,Jones2006}, but so far their utility has been limited. Since laser fields can be focused tighter and switched faster than magnetic fields, it is expected that OFRs could yield an MFR-like effect but with orders of magnitude better spatial and temporal control \cite{Yamazaki2010}.

Furthermore, OFRs are better suited for alkaline-earth-metal atoms, which have magnetically insensitive electronic ground states. The study of alkaline earth atoms is now a rich field, attracting attention for metrology \cite{Bloom2014}, quantum information \cite{Daley2008,Gorshkov2009}, and many-body physics \cite{Gorshkov2010}. Quantum degenerate gases of these atoms have also been realized \cite{Takasu2003,Kraft2009,Stellmer2013}. Many-body physics has been demonstrated in strontium lattice clocks \cite{Martin2013}, and it has been shown that controlling many-body interactions in gases of alkaline earth metals could lead to better clock accuracy \cite{Swallows2011}. Without MFRs to facilitate the same many-body control enjoyed by alkali-metal experiments, OFRs have been suggested as an alternative \cite{Ciuryo2005}.

OFRs have been the focus of several experiments. These resonances have been observed in alkali gases \cite{Fatemi2000,Theis2004} and in alkaline-earth-like atoms \cite{Enomoto2008}. P-wave OFRs have been reported \cite{Yamazaki2012}, and OFRs have been successfully applied to induce thermalization in Sr gases \cite{Blatt2011} and manipulate the condensate dynamics of a Sr Bose-Einstein condensate \cite{Yan2013a,Yan2013b}. The theory used to describe these experiments was based on a quantum defect treatment by Bohn and Julienne, who used an isolated resonance approximation to derive the optically modified scattering length \cite{Bohn1999}.

Although the isolated resonance theory has been successful in describing some observations of OFRs, it fails to explain OFR behavior in the large detuning regimes that are critical to a proposal for practically applying these resonances \cite{Ciuryo2005}. Attempts to experimentally realize this proposal did not succeed \cite{Blatt2011}; therefore, the limited theoretical understanding of OFRs has hindered experimental progress. To broaden the theoretical understanding of OFRs, we perform the most complete theoretical analysis of this effect to date. To this end, we treat OFRs with a numerical coupled channel method, which has been highly successful for treating MFRs \cite{Kohler2006,Chin2010}. Like the MFR theory, our coupled channel approach is capable of treating multiple interacting resonances without being restricted to the more limited isolated resonance approximation.  Consequently, this more general treatment allows us to study the range of validity of the isolated resonance approximation, and it also enables us to point out similarities and significant differences between OFRs and MFRs. Finally, we use the isolated resonance theory to derive experimentally useful formulas that can be used to understand real experimental OFR data.

\subsection{Basic Collision Theory}
\label{subsec:Basic}

In the context of cold-atom physics, a Feshbach resonance is a collisional resonance of two particles that is tunable by an external field. This is possible if a molecular bound state from an excited scattering channel (called the ``closed channel") couples to the free atom continuum of the ground state scattering channel (called the ``entrance channel" or the ``background channel"). Furthermore, the bound state energy is tunable by an external magnetic or electromagnetic field. The presence of this bound molecular state modifies the $s$-wave scattering length of the atoms, thereby changing the atomic interactions as the external field is tuned.

We emphasize that both MFRs and OFRs can be treated by the same scattering formalism, which accounts for the differences in their coupling and control mechanisms, as presented in the review by Chin {\it et al.}~\cite{Chin2010}.  A typical MFR is coupled to the entrance channel by internal short range spin-dependent couplings within the ground state manifold of Zeeman levels. MFRs are tuned by varying
an external magnetic field $B$ to move a molecular bound state across a collision threshold, creating a pole in the scattering $S$-matrix as a function of $B$. An OFR involves coupling a bound molecular state to two colliding atoms in their ground states using a photon from a laser, hereafter called the ``photoassociation laser" or ``PA laser.'' In this case, the coupling depends on both the PA laser detuning from a photoassociation resonance (the ``molecular detuning'') and the PA laser intensity. In contrast to MFRs, which are often based on molecular states that have very long lifetimes, spontaneous decay of the excited molecular state in an OFR introduces an appreciable linewidth to the molecular transition. Any population transferred to the excited state undergoes spontaneous decay, which translates to inelastic loss collisions that must be minimized in order to utilize an OFR. However, resonance decay does not necessarily prevent the application of OFRs since MFRs with 2-body decay channels ~\cite{Thompson2005b,Kohler2005,Naik2011} have proven experimentally useful ~\cite{Cornish2000,Donley2002,Papp2008,Trenkwalder2011,Kohlstall2012}.

\begin{figure}
  \centering
  \includegraphics[width=\linewidth]{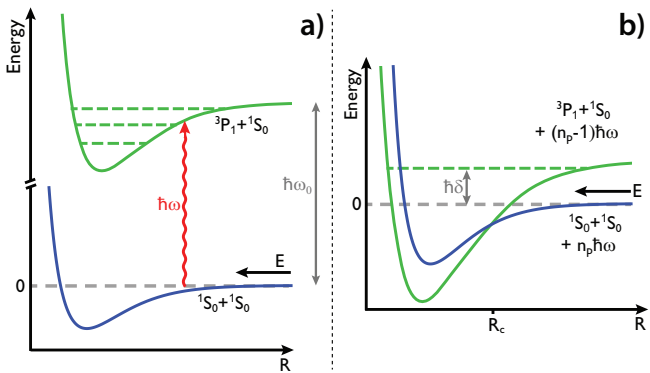}
  \caption{(Color online) a) The $^1S_0 + ^1\!S_0$ entrance channel of \Sr\ couples to a bound state of the $^3P_1 + ^1\!S_0$ closed channel via the PA laser field. The atomic transition is a 7.5 kHz intercombination line. Here $E$ is the collision energy, $\omega_0$ is the atomic resonance frequency, and $\omega$ is the laser frequency. b) In the dressed state picture, two free atoms in the entrance channel are brought into resonance with a molecular bound state. Here $\delta$ is the ``molecular detuning'' (defined in Section \ref{sec:isolated_res}), and $n_p$ is the photon number. The Condon radius $R_c$ is defined as the value of $R$ where the two potentials cross.}
  \label{fig:atoms_in_radiation_field}
\end{figure}

In the OFR studies presented here, we consider bosonic \Sr\ with the two ground state atoms providing the $^1S_0 + ^1\!\!S_0$ ground state entrance channel and the excited state $^1S_0 + ^3\!\!P_1$ providing the closed channels, schematically represented in Fig.~\ref{fig:atoms_in_radiation_field}. Since bosonic isotopes of alkaline earths have no nuclear spin, the \Sr\ resonance structure is considerably simpler than for atoms with hyperfine interactions, making it a good atom for an OFR experiment. The $^1S_0 \rightarrow ^3\!\!P_1$ atomic transition is an intercombination line with a natural linewidth of $\gamma_a = 2\pi \times$ 7.4 kHz. The narrowness of this transition means that all OFRs in \Sr\ are well resolved from the atomic line, decreasing the severity of off-resonant atomic light scattering.

To analyze the scattering of two colliding \Sr\ atoms in a light field, we calculate the scattering \emph{S}-matrix to determine the elastic and inelastic scattering cross sections. Since the \Sr\ ground state is completely spinless, and since current OFR experiments are typically performed at temperatures of a few $\mu$K or below, the scattering is described by an $s$-wave collision with a single nondegenerate entrance channel. Therefore, we will develop our theory for this experimentally simple OFR system, for which we only need a single $s$-wave $S$-matrix element $S(k)=e^{2i\eta(k)}$, represented in general by a complex energy-dependent phase shift $\eta(k)$. Here $k$ is defined via the collision velocity $\hbar k / \mu$, $\mu = m/2$ is the reduced mass, and $m$ is the mass of an \Sr\ atom. This phase shift in turn defines an energy-dependent scattering length $\alpha(k)$ as~\cite{Hutson2007,Idziaszek2010a,Quemener2012},
\begin{equation}
\label{eqn:scattering_length}
\alpha(k) = a(k) - i b(k) = -\frac{\tan \eta(k)}{k} = \frac{1}{i k} \frac{1 - S(k)}{1 + S(k)} \,.
\end{equation}
This expression is useful for small but nonvanishing collision energies, and the standard complex scattering length is the $k \to 0$ limit of this expression.  The elastic and inelastic loss cross sections are
\begin{align}
\sigma_{el} & = \frac{\pi g}{k^{2}} |1 - S(k)|^{2} =  8 \pi |\alpha(k)|^{2} f^{2}(k), \label{eqn:el_cross_section} \\
\sigma_{in} & = \frac{\pi g}{k^{2}} \left( 1 - |S(k)|^{2} \right) = \frac{8 \pi}{k} b(k) f(k). \label{eqn:in_cross_section}
\end{align}
Here $g$ is a collisional symmetry factor, which is equal to 2 for identical bosons (as assumed here). The function
\begin{equation}
\label{eqn:f_factor}
f(k) = \frac{1}{1 + k^{2} |\alpha(k)|^{2} + 2 k b(k)} \,
\end{equation}
approaches unity when $k|\alpha| \ll 1$ for all detunings. For a trapped gas of atoms, this limit occurs when $k_B T/\hbar \gamma \ll 1$, where $k_B$ is Boltzmann's constant and $T$ is the sample temperature. The elastic and inelastic collision rate coefficients are related to these cross sections as
\begin{align}
\label{eqn:rate_coefficients}
K_{el} (k) & = \frac{\hbar k}{\mu} \sigma_{el} (k)  \to 8\pi\frac{\hbar}{\mu} k|\alpha(k)|^2 \,\,\,\mathrm{as}\,\,\, k \to 0   \\
K_{in} (k) & = \frac{\hbar k}{\mu} \sigma_{in} (k) \to 8\pi\frac{\hbar}{\mu} b(k) \,\,\,\mathrm{as}\,\,\, k \to 0 \,. \label{eqn:loss_rate}
\end{align}
These general expressions are valid in the $s$-wave limit for OFRs and for decaying or non-decaying MFRs. A sum over higher partial waves is needed when these begin to contribute at higher $k$, and a thermal average of $K_{el}(k)$ and $K_{in}(k)$ is needed when comparing to experiment.

\section{Coupled Channels Formulation of Optical Feshbach Resonances}
\label{sec:CC}

\subsection{Background}
\label{subsec:CC_background}

The standard treatment for atomic collisions involving two or more internal states of the atoms is the coupled channels (CC) method.  Numerical models based on CC methods have been very successful in treating collisions and MFRs of ground state alkali-metal atoms~\cite{Chin2010,Kohler2006}. The CC method involves setting up a basis set representing the ``channels'' of the electronic, spin, and rotation degrees of freedom of the colliding atoms and then solving the matrix Schr{\"o}dinger equation for the amplitude of the radial motion in the interatomic separation coordinate $R$ for each of these channels.

In MFR theory, the channels represent the states of the atoms in a magnetic field for $R \rightarrow \infty$, the Born-Oppenheimer potentials characterize the $R$-dependent interactions, and spin coupling matrix elements are approximated by their atomic values. An external $B$ field shifts the energies of the atomic and molecular energy levels. In the case of OFRs, the channels represent the field dressed atoms, where the ground and excited states are coupled by the light field in the dipole approximation, and the $R$-dependent molecular interactions are represented by the ground and excited state Born-Oppenheimer potentials together with any non-adiabatic coupling between them. The OFR case has the added complication of including the spontaneous emission of light by excited state atoms or molecules.

To date, all cold atom OFRs have been treated by a resonant scattering formulation ~\cite{Thorsheim1987,Fedichev1996,Bohn1999}. The next section will discuss the approximation of treating each OFR as an isolated resonance. Here we concentrate on giving a full CC treatment ~\cite{Zimmerman1977,DeVries1978a,DeVries1978b,Mies1981} that includes the effect of multiple overlapping resonances without restricting the theory to treating isolated single resonances. This enables us to establish the conditions under which the isolated resonance approximation is valid.

We follow the field-dressed collision approach of Julienne~\cite{Julienne1982a,Julienne1982b}, which was applied to explain experiments on the collisional redistribution of light ~\cite{Julienne1984,Julienne1986}. To do this it is necessary to properly represent the three-dimensional (3D) nature of the collision and the role of atomic degeneracy. References~\cite{Julienne1982b,Napolitano1997} treat the exchange of multiple photons during a collision, by which one partial wave is coupled to higher partial waves through the intrinsically anisotropic nature of the interaction with light. Reference~\cite{Napolitano1997} adapts the CC dressed atom formalism to cold atom collisions in strong optical fields to explain the phenomena of optical shielding.

Three effects need to be incorporated within a CC theory to describe OFRs in the collision of $^1$S$_0$ Sr atoms in a light field tuned near the $^1S_0 \rightarrow ^3\!\!P_1$ line: (1) the field dressing of the collision, (2) the inherently 3D nature of the collision, with a space axis defined for the separated atoms by the PA laser polarization but with a rotating interatomic axis needed for the excited molecular bound states, and (3) the spontaneous emission while in the excited state.  If we make the approximation that the light field is weak, the total angular momentum $J$ is a good quantum number (the optical coupling matrix element remains small compared to the spacing of rotational levels in the excited state). In this case Refs.~\cite{Julienne1984,Julienne1986} showed that six CCs are needed to describe optically coupled $^1$S$_0+ ^1$S$_0 \to ^1$S$_0+ ^1$P$_1$ collisions.  The same is true when we replace $^1$P$_1$ with $^3$P$_1$. One set of channels represents the ground state collision with partial wave $\ell=J$ and $n_p$ photons at an angular frequency $\omega$. Another set represents the excited $0_u$ and $1_u$ molecular states with $n_p-1$ photons at frequency $\omega$ and respective projection $\Omega=$ 0 and 1 of electronic angular momentum $j=1$ on the interatomic axis. These excited state channels have total angular momentum $J_e=J-1$ (two channels), $J$ (one channel), and $J+1$ (two channels).  In the special case of $s$-wave collisions ($J=0$) of cold atoms, only three channels are needed, representing the ground state and the $0_u$ and $1_u$ states with $J=1$. Finally, spontaneous emission from the excited molecular state can be included with a complex potential ~\cite{Napolitano1994}, with a caveat that the imaginary decay part of the potential has to be turned off when the atoms are far apart in the free atom limit.

We assume that the free atoms are weakly dressed---that is, the PA laser with photon energy $\hbar\omega$ is detuned from the atomic excitation energy $\hbar\omega_0$ by a large amount compared to the optical coupling strength
\begin{equation}
V_\mathrm{opt}=\left (2\pi I/c\right )^{1/2} d \,,
\label{eqn:Vopt}
\end{equation}
where $I$ is PA laser intensity, $c$ the speed of light, and $d$ is a molecular transition dipole matrix element~\cite{Julienne1986}. However, the short range molecular states can be strongly dressed, so that the peak of an on-resonant PA line at $\hbar\omega_n$, where $n$ is the molecular vibrational level, can be power broadened.  The rotational quantum number $J_e$ will remain a good approximate quantum number as long as $V_\mathrm{opt}$ remains small compared to the rotational constant $B_n$ of level $n$. (The separations of the $J=0$ and 2 levels from the $J=1$ level are $2B_n$ and $4B_n$ respectively.)

\subsection{Formulation for $^{88}$Sr}
\label{subsec:CC_formulation}

We include in our treatment here the minimal number of three channels needed to get a basic description of near-threshold $s$-wave OFRs.   This minimal treatment could be written in either of two basis sets representing the electronic, spin, rotational, and photon  degrees of freedom.  One basis set for the molecular degrees of freedom is the Hund's case (c) basis represented as $|\Omega_sJM\rangle_c$, where the projection of electronic plus spin angular momentum $j$ on the rotating body-fixed axis is $\Omega$, $s$ represents the {\it gerade} or {\it ungerade} inversion symmetry of electronic coordinates, and $M$ is the projection of total angular momentum $J$ on a space-fixed axis.  The other molecular basis is the Hund's case (e) asymptotic basis of Refs.~\cite{Julienne1982a,Julienne1986} represented as $|j_s\ell JM\rangle_e$, where $j_s=0$ or 1 represents the separated atoms in the respective $^{1}S_{0} + ^{1} \! S_{0}$ and $^{1}S_{0} + ^{3} \! P_{1}$  channels with partial wave $\ell$, coupled to total angular momentum $J$ and projection $M$.  The subscript $s$ on $j$ indicates that the electronic wavefunction is symmetrized with respect to the exchange of electronic coordinates.  Table~\ref{tab:basis} shows the three basis functions for a dressed CC calculation in either representation.    The transformation between the molecular and asymptotic representations is (see, for example, Eq.~(36) of Ref.~\cite{Julienne1982a}):
\begin{eqnarray}
|0_uJM\rangle_c &=& \left (\frac{J}{2J+1}\right)^{1/2} |1_u,J-1,JM\rangle_e \nonumber \\ & &- \left (\frac{J+1}{2J+1}\right)^{1/2} |1_u,J+1,JM\rangle_e \label{eqn:c-e1} \\
|1_uJM\rangle_c &=& \left (\frac{J+1}{2J+1}\right)^{1/2} |1_u,J-1,JM\rangle_e \nonumber \\ & & + \left (\frac{J}{2J+1}\right)^{1/2} |1_u,J+1,JM\rangle_e \,. \label{eqn:c-e2}
\end{eqnarray}
\begin{table}
\caption{Minimal CC basis sets in the Hund's case $b=$ (c) and (e) representations, where $\sigma=0,\pm1$ represents the polarization of the light with $n_p$ photons of frequency $\omega$ that couples the ground and excited states.}
\label{tab:basis}
\begin{tabular}{ccc}
\hline\hline
Channel & Case (c): $|\Omega_sJM\rangle|n_p\omega\sigma\rangle$ & Case (e): $|j\ell JM\rangle|n_p\omega\sigma\rangle$ \\
\hline
$|1\rangle_b$ & $|0_g00\rangle_c|n_p\omega\sigma\rangle$ & $|0_g000\rangle_e|n_p\omega\sigma\rangle$ \\
\rule{0pt}{2ex}
$|2\rangle_b$ & $|0_u1\sigma\rangle_c|n_p-1,\omega\sigma\rangle$ & $|1_u01\sigma\rangle_e|n_p-1,\omega\sigma\rangle$ \\
$|3\rangle_b$ & $|1_u1\sigma\rangle_c|n_p-1\omega\sigma\rangle$& $|1_u21\sigma\rangle_e|n_p-1\omega\sigma\rangle$ \\
\hline
\end{tabular}
\end{table}

Using the CC expansion of the full wavefunction at total energy $E$,
\begin{equation}
  \Psi(R,E) = \sum_{i=1}^3 |i\rangle_b F_{i,b}(R,E)/R
  \label{eqn:CC_Psi}
\end{equation}
where $F_{i,b}$ represents the amplitude of the wavefunction projected on the basis function $| i \rangle_b$ for Hund's case b = (c) or (e). The CC matrix Schr{\"o}dinger equation for the $s$-wave collision of the two atoms in a (moderately) weak light field is
\begin{equation}
  \frac{\partial^2{\Psi}}{\partial R^2} + \frac{2\mu}{\hbar^2} \left (E\cdot \bf{I} - \bf{V}(R)  \right )\Psi = 0
  \label{eqn:CC}
\end{equation}
where $\bf{I}$ is the identity matrix and the potential matrix $\bf{V}$ describes the diagonal and off-diagonal matrix elements of the collisional and optical interactions. Either the $b=$ (c) or (e) representations (Table \ref{tab:basis}) of the excited state could be used to set up the expansion and $\bf{V}$ matrix in Eqs.~(\ref{eqn:CC_Psi}) and~(\ref{eqn:CC}). Our numerical calculations use the Hund's case (e) basis, for which the matrix elements are given in Table I of Ref.~\cite{Julienne1986}, and quoted in the supplemental online material for Ref.~\cite{Blatt2011}:
\begin{equation}
\label{eqn:CC_V}
\bf{V} = \left( \begin{array}{ccc}
V_g & V_\mathrm{opt} & 0  \\
 V_\mathrm{opt} &  \frac{1}{3}(V_{0u}+2V_{1u}) & \frac{\sqrt{2}}{3}(V_{1u}-V_{0u}) \\
0 &  \frac{\sqrt{2}}{3}(V_{1u}-V_{0u}) &  \frac{1}{3}(2V_{0u}+V_{1u})+6 V_\mathrm{cen}
 \end{array} \right ) \, ,
\end{equation}
where the $6 V_\mathrm{cen}$ term represents the $d$-wave centrifugal potential with $V_\mathrm{cen}=\hbar^2/(2\mu R^2)$. Here $V_g(R)$, $V_{0u}(R)$, and $V_{1u}(R)$ represent the ground state and $0_u$ and $1_u$ excited state BO potentials, each of which we model as a Lennard-Jones potential plus an additional long range term:
\begin{eqnarray}
V_g(R) &=& \left ( \left(\frac{R_{0,g}}{R}\right)^6 - 1 \right ) \frac{C_{6,g}}{R^6} -\frac{C_{8,g}}{R^8} +V_{g\infty}  \label{eqn:Vg}\\
V_{0u}(R) &=&  \left (  \left(\frac{R_{0,0u}}{R}\right)^6 - 1 \right ) \frac{C_{6,0u}}{R^6}-\frac{C_{3,0u}}{R^3} +V_{u\infty} \label{eqn:V0u} \\
V_{1u}(R) &=&  \left (  \left(\frac{R_{0,1u}}{R}\right)^6 - 1 \right ) \frac{C_{6,1u}}{R^6}+\frac{C_{3,1u}}{R^3} +V_{u\infty} \label{eqn:V1u} \,,
\end{eqnarray}
The $V_{s\infty}$ terms give the asymptotic values of the potentials as $R \to \infty$, as explained below.  We use the excited state potential parameters from Zelevinsky {\it et al.}~\cite{Zelevinsky2006,footnote1}.  The ground state $C_{6,g}$ and $C_{8,g}$ parameters come from Ref.~\cite{Porsev2006}, and $R_{0,g}$ was optimized to reproduce the measured bound state binding energies of Ref.~\cite{Escobar2008} to within 0.4\% ~\cite{footnote2}. $V_g(R)$ has an $s$-wave scattering length of $-1.4$ a$_0$, consistent with that reported in Ref.~\cite{Escobar2008}.

The optical coupling matrix element in Eq.~(\ref{eqn:CC_V}) is given by Eq.~(\ref{eqn:Vopt}) in the dipole approximation, where we neglect retardation (that is, the phase difference between the optical fields separated by distance $R \ll \lambda$, where $\lambda=2\pi c/\omega$ is the wavelength of the excitation light). Thus, since we use the symmetrized $g$ and $u$ electronic states, $d=\sqrt{2} d_a$, where the atomic transition dipole $d_a= 0.08682$ atomic units (1 a.u. $=$ $ea_0$, where $e$ is the electron charge and $a_0$ is the Bohr radius), corresponding to an atomic $^3$P$_1$ lifetime of 21.46 $\mu$s or linewidth of $\gamma_a$ $=$ $2 \pi \times$ 7.416 kHz.  Thus, introducing units into Eq.~(\ref{eqn:Vopt}),
\begin{equation}
    V_\mathrm{opt}/h = 24.83 \, \mathrm{MHz} \, \times \, d_a \sqrt{I/\mathrm{(1 W/cm}^2)} \,,
\end{equation}
where $d_a$ is in atomic units. The optical coupling in $\bf{V}$ conforms to the case (e) selection rule that $\Delta \ell =0, \Delta m_\ell =0$ (it is only the electronic $j$ quantum number that changes).  This coupling is also independent of light polarization $\sigma$ for this transition. Using Eq.~(\ref{eqn:CC_V}), there will be an asymptotic light shift
\begin{equation}
  V_\infty=  \frac{\hbar(\omega_0-\omega)}{2} \left ( \sqrt{\left ( \frac{2 V_\mathrm{opt}}{\hbar(\omega_0-\omega)}\right )^2+1}-1 \right ) ,
\end{equation}
which is negative for the ground state and positive for the excited state.  Thus, taking $V_{g\infty}=V_\infty$ and $V_{u\infty}=\hbar(\omega_0-\omega)+V_\infty$ in Eqs.~(\ref{eqn:Vg})-(\ref{eqn:V1u}) ensures that when $\bf{V}$ is diagonalized the lowest energy eigenvalue at large $R$ for the field-dressed ground state is zero.  With this definition of the zero of energy, the total energy $E$ in the CC Schr{\"o}dinger equation~(\ref{eqn:CC}) represents the relative collision kinetic energy $\hbar^{2} k^{2}/2 \mu$ of the dressed ground state atoms, and $E = \hbar^{2} k^{2}/2 \mu \to 0$ at the collision threshold.

The matrix $\bf{V}$ in Eq.~(\ref{eqn:CC_V}) could be transformed to the molecular case (c) representation by transforming the $2 \times 2$ excited state block using the (c) to (e) transformation matrix used in Eqs.~(\ref{eqn:c-e1}) and~(\ref{eqn:c-e2}).  This would give the diagonal $J=1$ $0_u$ and $1_u$ potentials given in Eqs.~(1) and (2) of Zelevinsky {\it et al.}~\cite{Zelevinsky2006} and generate the body-frame Coriolis coupling term between these two states.  The optical coupling in the case (c) molecular basis is different from that in the asymptotic case (e) basis.  Using the transformations in Eqs.~(\ref{eqn:c-e1}) and~(\ref{eqn:c-e2}) shows that the optical couplings matrix elements between the ground $J=0$ $0_g$ state and the respective excited $J=1$ $0_u$ and $1_u$ states are determined from Eq.~(\ref{eqn:Vopt}) with the molecular dipole matrix elements
\begin{eqnarray}
 d_{0u} &=& \sqrt{1/3} \sqrt{2}d_a \,, \label{eqn:d0u} \\
 d_{1u} &=& \sqrt{2/3} \sqrt{2}d_a \,. \label{eqn:d1u}
\end{eqnarray}
The $\sqrt{2}$ is the same homonuclear $ g \to u$ enhancement factor that affects the case (e) matrix element.  The other factor corresponds to the usual H{\"o}nl-London factor for R-branch ($J \to J+1$) molecular transitions.

Treating an OFR requires that we include the decay from the excited state, which has a molecular decay rate $\gamma$. Our calculations assume $\gamma=\gamma_m$, where we define $\gamma_m=2\gamma_a$. This rate $\gamma_m$ is the rate of spontaneous emission from the excited molecular state in the long-range non-retarded dipole approximation. A nonzero value of $\gamma-\gamma_m$ would be due to other processes that induce decay of the excited state or change the emission rate from its long-range non-retarded dipolar value. While Bohn and Julienne~\cite{Bohn1999} introduced artificial channels to simulate excited state decay, a simpler method is to introduce an imaginary term in the excited state potentials in Eqs.~(\ref{eqn:V0u}) and (\ref{eqn:V1u}), replacing $V_{ju}$, $j=0,1$, with:
\begin{equation}
  V_{ju} -i\frac{\hbar \gamma}{2}  \left ( 1+e^{\beta (R-R_\mathrm{cut})} \right )^{-1} ,
  \label{eqn:ImV}
\end{equation}
where $\gamma$ is an $R$-independent constant. The function in parenthesis ensures that molecular decay turns off at large distances when $R$ exceeds the arbitrary cutoff radius $R_\mathrm{cut}$ by an amount large compared to the length $1/\beta$. Furthermore, this function ensures that the full molecular decay rate turns on at small distances where $R_\mathrm{cut} - R$ is appreciably less than $1/\beta$. The constant $\beta$ parametrizes the distance over which molecular decay becomes appreciable.

When the coupled equations are solved with this complex potential in Eqs.~(\ref{eqn:V0u}) and (\ref{eqn:V1u}), the $S$-matrix is non-unitary, and $1 - |S(k)|^2$ in Eq.~(\ref{eqn:in_cross_section}) represents loss of ground state atoms due to molecular excitation followed by excited state decay. We assume that every spontaneous emission event represented by the imaginary term in $V_{ju}$ results in hot atom or molecular products that are lost from the trap. Our numerical studies show that this assumption is good for all the excited levels except the state nearest in energy to the atomic resonance (Section \ref{subsec:CC_Results}). We calculate that 60\% of the emission from this state does not result in loss from a 10 $\mu$K trap~\cite{Zelevinsky2006,Reinaudi2012}. The cutoff ensures that there is no spurious excited state decay associated with the asymptotically dressed atoms. We find that in the core of a PA line, out to molecular detunings of several hundred line widths from molecular resonance, the loss associated with the imaginary part of the scattering length is not sensitive to the value chosen for $R_\mathrm{cut}$, as long as it is significantly outside the outer turning point of classical motion for the excited state vibrational level. Furthermore, the real part of the scattering length is completely insensitive to the choice of $R_\mathrm{cut}$. We typically choose $R_\mathrm{cut}=$ 500 a$_0$ and $\beta=$ 0.05 a$_0^{-1}$. We find that our numerical calculations were insensitive to the choice of $\beta$.

 \subsection{Approximations and limitations}
 \label{subsec:CC_approx}

This three-channel model makes several approximations, but is capable of representing the essential qualitative and semi-quantitative effects associated with OFRs in the weak to moderate field regime. Our model only includes the minimal number of partial waves needed to represent the change in scattering length and molecular losses due to the OFR. It leaves out the coupling of the excited $J=1$ levels to the ground state $d$-waves as well as coupling to the doubly excited states associated with the $^3$P$_1$ $+$  $^3$P$_1$ separated atom limit. This means that the light shifts calculated from the three-channel model will not be accurate, since ground state $d$-waves are known to contribute to PA light shifts~\cite{Simoni2002,Ciurylo2006}, and doubly excited states may contribute also.  Furthermore, the effect of field-dressing in modifying the ground state threshold elastic scattering of partial waves with $\ell >0$ is not included.  This modification is due to field dressing that brings in $1/R^3$ terms in the long range potential that will affect the $d$-wave collisions of like bosons or the $p$-wave collisions of like fermions or unlike species.  Note that in our three-channel treatment, the field dressed $s$-wave interactions have the correct property that they have no long-range $1/R^3$ component, since such variation vanishes in the $V_{2,2}$ matrix element of Eq.~(\ref{eqn:CC_V}) (if we had attempted only a two-channel field dressed treatment, the presence of the single $0_u$ excited state potential would have introduced a spurious $1/R^3$ term in the ground state dressed $s$-wave potential).

It would be straightforward to introduce higher partial waves and strong field dressing into the calculation, using the formalism of Napolitano {\it et al.}~\cite{Napolitano1997}.  This formalism uses the ``uncoupled" asymptotic basis $|j m_j \ell m_\ell\rangle$, which is better for treating strong field dressing than the ``coupled'' $|j \ell JM\rangle$ basis we use here. The subtle effects of retardation, switching off the dipole approximation, and including the weak coupling to the {\it gerade} states as $R$ increases~\cite{Takasu2012} should be taken into account in a more complete theory. We do not perform a time-domain analysis, so we cannot reproduce the transient OFR dynamics \cite{Naidon2008} observed in Ref. \cite{Yan2013b}. Furthermore, a full treatment of excited state spontaneous emission during a collision is beyond the scope of CC methods, and would require treatment by stochastic Schr{\"o}dinger equation methods (density matrix methods are computationally intractable)~\cite{Suominen1994,Suominen1998a}. Last, as we show in the next section, our analysis with a cutoff of the long range decay is sufficient for treating OFRs for molecular detunings that are of order 100 line widths (or less) from the center of a PA line.

\subsection{Coupled Channels Results}
\label{subsec:CC_Results}

The CC calculations to solve Eq.~(\ref{eqn:CC}) were carried out using the standard renormalized Numerov method~\cite{Johnson1977} using complex variables so as to represent the effect of the complex potential in the excited state channels.  A step-doubling algorithm was employed to optimize the number of steps needed as $R$ increases between the short and long range regions.  The single $S$-matrix element $S(E,I,\omega)$ for the dressed ground state $s$-wave was extracted from the log derivative of the single open channel solution $F_{1,e}(R)$ of the three channel propagated wavefunction of Eq.~(\ref{eqn:CC_Psi}) at a suitable large asymptotic value of $R$.   Using Eq.~(\ref{eqn:scattering_length}) then gives the complex energy-dependent scattering length $\alpha(k,I,\omega)$, which then gives the elastic and inelastic rate coefficients $K_{el}$ and $K_{in}$ (Eqns. \ref{eqn:rate_coefficients} and \ref{eqn:loss_rate}).

\begin{figure}
  \centering
  \includegraphics[width=\linewidth]{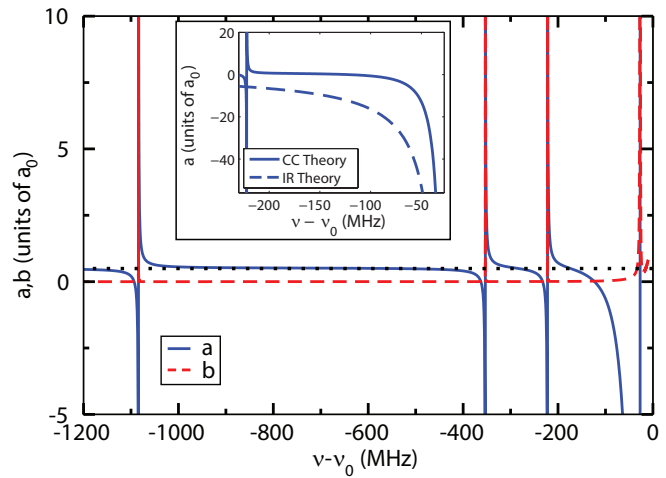}
  \caption{(Color online) Real (solid line) and imaginary (dashed line) parts of the complex scattering length $\alpha=a-ib$. Here the detuning from atomic resonance $\nu-\nu_0$ (the ``atomic detuning'') is measured with respect to the $^{88}$Sr intercombination line transition at $\nu_0$. Also, $E/k_B=4$ $\mu$K and $I=10$ W/cm$^2$.  The dotted line shows the background $a_{bg}$ for $E/k_B=4$ $\mu$K and $I=0$. Inset: A close up of the off-resonant behavior of the $n = -2$ OFR (solid line). Also plotted is the scattering length predicted by treating the $n = -2$ OFR as an isolated resonance (dashed line).}
  \label{fig:CC_abscan}
\end{figure}
Fig.~\ref{fig:CC_abscan} shows the real and imaginary parts of $\alpha(k)$ as the PA laser frequency $\nu = \omega/2\pi$ is detuned from atomic resonance at $\nu_0=\omega_0/2\pi$ (where $\nu - \nu_0$ is the ``atomic detuning'').  This particular example was taken for a PA laser intensity of 10 W/cm$^2$ and a relative collision kinetic energy of $E/k_B=4$ $\mu$K. Here the background $a_\mathrm{bg}=$ 0.495 a$_0$ at $E/k_B=4$ $\mu$K differs from the background value -1.4 a$_0$ in the limit of $E=0$ due to the energy dependence of $a_\mathrm{bg}(k)$. The calculations were carried out for atomic detunings larger in magnitude than -20 MHz to avoid strong field-dressing effects at atomic detunings near resonance (the optical coupling matrix element $V_\mathrm{opt}/h=0.84$ MHz for this $I$).  The decay rate was taken to be $\gamma =\gamma_m=2\gamma_a= 2 \pi \times$ 0.014833 MHz. The figure shows a series of four OFRs in this region. These four resonances correspond to the previously observed~\cite{Zelevinsky2006} $n=$ -2, -3, and -4 members of the $0_u$ $J=1$ series at binding energies $E_n/h =$ 24 MHz, 222 MHz, and 1084 MHz and a single $n=$ -1 member of the $1_u$ $J=1$ series at 353 MHz. Here $n<0$ counts bound states down from the last level (of a given $0_u$ or $1_u$ symmetry) designated as $n=-1$. The scattering lengths show a series of overlapping resonances that cause a large change in scattering length near the poles of the resonances but return to $a_\mathrm{bg}$ between resonances. The fact that the stronger $n = -2$ resonance returns to its background value near the $n = -3$ line (Fig. \ref{fig:CC_abscan} inset) illustrates an important general feature of a vibrational sequence of OFRs: interfering resonances cause the scattering length to return to its background value in between resonances. Even the presence of a neighboring OFR that is comparatively weak will diminish the off-resonant magnitude of a stronger OFR (Fig. \ref{fig:CC_abscan} inset). This property imposes a constraint on OFR experiments, namely that molecular detunings cannot be so large as to be comparable to the frequency separation between the resonance of interest and the nearest resonance. In contrast, MFRs arising from neighboring spin-channel resonances interfere with one another in a manner that is qualitatively different than for a vibrational series (see Section \ref{subsec:OFR_MFR_multires}).

\begin{figure}
  \centering
  \includegraphics[width=\linewidth]{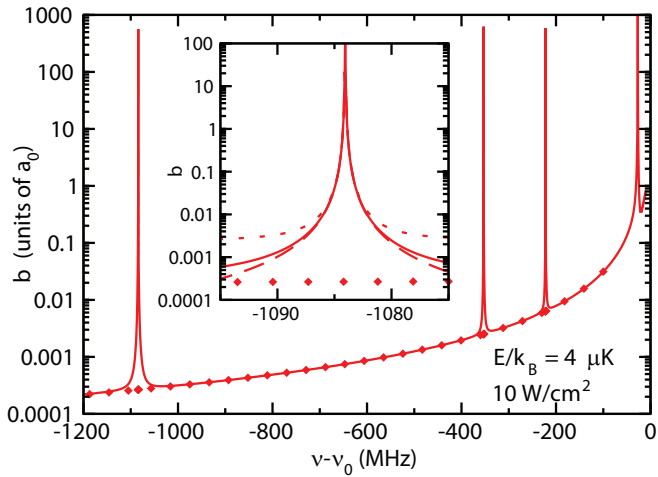}
  \caption{(Color online) Imaginary part of $\alpha = a - i b$ from Fig.~\ref{fig:CC_abscan} shown on a log scale.  The diamonds show a $1/(\nu-\nu_0)^2$ scaling.  The inset shows an expanded view of the $0_u$ $n=-4$ resonance near $-1084$ MHz.  The dashed, solid, and dotted lines show the results for $R_\mathrm{cut}=$ 200 a$_0$, 500 a$_0$, and 1000 a$_0$ respectively.  Near the peak of the resonance, $b$ is independent of $R_\mathrm{cut}$.}
  \label{fig:CC_bscan}
\end{figure}
The imaginary part $b$ of the scattering length that gives the loss rate coefficient, Eq.~(\ref{eqn:loss_rate}),  shows a series of sharp spikes near the poles of the resonances in Fig.~\ref{fig:CC_abscan}, and shows a value very near zero on the linear scale of the figure.  Figure~\ref{fig:CC_bscan} provides a better way to illustrate the basic features of $b$ by showing it on a log scale.  Here the ``background'' on which the poles sit varies as $1/(\nu-\nu_0)^2$, with this functional form indicated by the diamonds on the figure. Furthermore, far detuned from a molecular resonance, the magnitude of this background is found to scale linearly with $R_\mathrm{cut}$ as $R_\mathrm{cut}$ increases. This is because away from resonance, most of the loss of flux in the collision due to the presence of a complex potential comes from the long range region, where decay should not be counted as loss, since it merely represents atomic light scattering that returns an atom to its ground state.  Thus, loss is overcounted by use of a complex potential if $R_\mathrm{cut}$ is too large.

In the core of the line spanning molecular detunings of $100 \gamma_m$, $b$ and $K_{in}$ are independent of  $R_\mathrm{cut}$. For example at a molecular detuning of $100 \gamma_m$ (inset to Fig.~\ref{fig:CC_bscan}) the values of $b$ calculated with $R_\mathrm{cut}=$ 200 a$_0$ or 500 a$_0$ differ by less than \%. The difference grows to 10\% if $R_\mathrm{cut}=$ 1000 a$_0$.  The difference with  $R_\mathrm{cut}=$ 200 a$_0$ or 500 a$_0$ only grows to 10\% when the molecular detuning is over 250 line widths.  Consequently, if $R_\mathrm{cut}$ is selected to have a small enough ``physical'' value where spontaneous decay for $R<R_\mathrm{cut}$ represents actual loss of atoms, the loss calculated for molecular detunings up to a few hundred line widths are meaningful and not sensitive to the choice of $R_\mathrm{cut}$.  In any case, the scattering length $a$ given by the real part of $\alpha$ is completely insensitive at all detunings to the choice of $R_\mathrm{cut}$. Comparing our CC theory to experimental data taken at small molecular detunings, we are able to reproduce the resonance strengths measured in Refs. \cite{Blatt2011} and \cite{Yan2013a}. However, the atom loss rate of Ref. \cite{Yan2013a} is measured at a molecular detuning large enough for our theory to be sensitive to $R_\mathrm{cut}$; therefore, our theory is not designed to reproduce this rate.

Within the inherent limitations of our approximations that we have outlined above, we expect our CC calculations to give the correct change in scattering length for all detunings and the atom loss rate coefficient for at least 100 line widths from the peak of a molecular resonance.  Consequently, since the resonances are spaced by many thousands of line widths apart, we turn our attention in the next section to understanding the theory for single isolated OFRs for molecular detunings in the vicinity of a photoassociation resonance.

\section{Isolated Resonance Theory of Optical Feshbach Resonances}
\label{sec:isolated_res}

The OFR features in Figs.~\ref{fig:CC_abscan} and~\ref{fig:CC_bscan} tend to be well-isolated from one another and thus can be described quite successfully by theory designed to treat an isolated single resonance situated on a background.  Isolated resonance theory has been used for cold atom OFRs since they were first proposed ~\cite{Thorsheim1987,Napolitano1994,Fedichev1996,Bohn1997,Bohn1999}.  This theory successfully explained alkali-metal atom PA spectra with hyperfine structure (in good agreement with experiment~\cite{Napolitano1994,Tiesinga2005}), and it also explained the saturation of PA lines at high intensity~\cite{McKenzie2002,Prodan2003,Simoni2002}. In fact, both OFRs and MFRs can be treated by the same resonance scattering formalism when the possibility of decay of the closed channel resonance state is taken into account~\cite{Chin2010}.

The isolated resonance approximation assumes that each molecular bound state is far from the other molecular states in the closed channel and can be described by a strength parameter that is local to the resonance---that is, independent of energy and molecular detuning in the vicinity of the resonance. Bohn and Julienne give a general resonance scattering treatment for an OFR based on quantum defect theory~\cite{Bohn1999}.  They derive a general expression for the $S$-matrix element $S(k)$ for a single $s$-wave entrance channel coupled to an isolated resonance scattering bound state, including a decay rate $\gamma$ to exit channels that lead to atom loss. The elastic scattering $S$-matrix element $S(k)$ (equivalent to Eq.~(3.13) of Ref.~\cite{Bohn1999}) for an isolated decaying resonance is:
\begin{equation}
\label{eqn:bohn_s_matrix}
S(k) = \left( 1 - \frac{i \hbar\Gamma(k)}{ E - E_\mathrm{res}+ i\frac{1}{2}\hbar [\gamma + \Gamma(k)]} \right) e^{2 i \eta_{bg}(k)} \,.
\end{equation}
where
\begin{equation}
\label{eqn:Eres}
 E_\mathrm{res} =\hbar ( \omega_{n}+s_n I - \omega)  = -\hbar\delta
\end{equation}
is the energy of the field-dressed molecular resonance level $n$. Its ``bare'' location at $\hbar \omega_n=\hbar \omega_a-E_n$  is shifted by an intensity-dependent shift $\hbar s_n I$, where $E_n$ is the binding energy with respect to the excited separated atom limit when $I=0$. We define the molecular detuning $\delta$ so it is negative for red detuning, in which case a resonance peak occurs when $E=E_\mathrm{res}$.

We assume low power, in which case the shift varies linearly with intensity. The coefficient $s_n$ can be either positive or negative~~\cite{Bohn1999,Ciurylo2006}, where a positive value corresponds to a shift of the resonance peak closer to the atomic line. The closed channel resonance bound state is coupled to the entrance channel by the stimulated decay rate,
\begin{equation}
\label{eqn:Gamma}
  \Gamma(k) =  \frac{2\pi}{\hbar} |\langle n |V_\mathrm{opt}|E\rangle |^2 \,.
\end{equation}
Here, $|n\rangle$ represents the excited bound state, which in general would be a mixture of the two $|2\rangle$ and $|3\rangle$ excited case (c) states in Table I.  In practice $|n\rangle$ would be well-approximated by a single $0_u$ or $1_u$ $J=1$ vibrational state.   The ground state scattering wavefunction $|E\rangle$ is assumed to be energy normalized \cite{Julienne2009}, so that
\begin{equation}
    F_1(R,E) \to \left ( \frac{2\mu}{\pi \hbar^2 k} \right )^{1/2} \sin(kR +\eta_\mathrm{bg}) \,\,\mathrm{as} \,\, R \to \infty \,,
\end{equation}
where the phase shift $\eta_\mathrm{bg}$ is related to the scattering length $a_\mathrm{bg}$ in the $k \to 0$ threshold limit as $\eta_\mathrm{bg} = -ka_\mathrm{bg}$.

We emphasize that the form of the expression in Eq.~(\ref{eqn:bohn_s_matrix}) is completely general for any isolated threshold resonance and applies equally well for MFRs and OFRs, if the terms are identified properly.    The Fermi golden rule width $\Gamma(k)$ expresses the strength of the resonance pole term with a tunable denominator.  When $\Gamma(k)=0$, there is no Feshbach resonance, and we recover the standard expression $S(k)=e^{2i\eta_\mathrm{bg}}$ for the uncoupled entrance channel. The expression in Eq.~(\ref{eqn:Gamma}) ensures that $\Gamma(k)$ follows the standard threshold law, and thus for an $s$-wave entrance channel, $\Gamma(k) \propto k$. For non-decaying resonances, $\gamma=0$, and the imaginary term in the denominator vanishes as $k \to 0$.

\section{Comparison Between Optical Feshbach Resonances and Magnetic Feshbach Resonances}

\subsection{Isolated Resonance Theory}
\label{subsec:OFR_MFR_isores}

The resonance length formalism summarized in Section II.A.3 of Ref.~\cite{Chin2010} shows how to relate MFR and OFR resonance strengths and compare OFRs to MFRs in a unified approach.   All we need to note is that in the case of an isolated MFR, the threshold width $\Gamma(k)$ in Eq.~(\ref{eqn:bohn_s_matrix}) is given by an expression similar to Eq.~(\ref{eqn:Gamma}), except that $V_\mathrm{opt}$ needs to be replaced with an appropriate internal spin-dependent Hamiltonian~\cite{Kohler2006,Chin2010}.  Furthermore, $E_\mathrm{res}$ would be replaced with a $B$-dependent tuning and shift~\cite{Julienne2006,Chin2010,Jachymski2013}, $E_\mathrm{res} = \delta\mu (B-B_c)+E_\mathrm{shift}$, where $B$ represents magnetic field, $B_c$ is the field where the bare resonance level crosses threshold, $\delta \mu$ represents the difference between the sum of the magnetic moments of the two atoms and the magnetic moment of the bare resonance level, and $E_\mathrm{shift}$ represents an energy-dependent shift term.

The threshold law for $s$-wave collisions shows that as $k \to 0$ the quantity $\hbar \Gamma(k)/k$ (for either an OFR or the MFR analog) becomes a $k$-independent constant having the units of length times energy.  Thus, for either an MFR or an OFR, we can decompose $\Gamma(k)/k$ into a product of a length factor $L_\mathrm{r}$ and an energy $E_\mathrm{r}$,

\begin{equation}
  \frac{\hbar \Gamma(k)}{2k} = L_\mathrm{r} E_\mathrm{r}\,,
  \label{eqn:res_constant}
\end{equation}
Since only the $L_\mathrm{r} E_\mathrm{r}$ product is significant, we are free to choose either the length $L_\mathrm{r}$ or the energy $E_\mathrm{r}$ factor to yield a convenient expression for the scattering length.

In the case of non-decaying MFRs, it is conventional to choose $L_\mathrm{r}= a_\mathrm{bg}$. The $E_\mathrm{r}$ factor is typically written as  $ \delta\mu \, \Delta$, thereby defining the magnetic ``width'' $\Delta$ of the MFR:
\begin{equation}
\label{eqn:Gamma_MFR}
\frac{ \hbar \Gamma_\mathrm{MFR}(k)}{2k} = a_\mathrm{bg} ( \delta\mu \, \Delta ) \,.
 \end{equation}
Here the subscript ``MFR'' indicates the type of resonance. When this form is substituted in Eq.~(\ref{eqn:bohn_s_matrix}), $\gamma$ is set equal to zero, and the $k \to 0$ limit is taken with the shift term in Refs.~\cite{Chin2010,Julienne2006}, Eq. (\ref{eqn:scattering_length}) reduces to the standard expression for an MFR,
\begin{equation}
\label{eqn:a_MFR}
  a(B) = a_\mathrm{bg} - a_\mathrm{bg} \frac{\Delta}{B-B_0} \,,
\end{equation}
where $B_0 = B_c-E_{\mathrm{shift}}/\delta\mu$ is the pole position. In the case $\gamma \neq 0$, this procedure would give the complex scattering length for a decaying MFR \cite{Chin2010,Naik2011}.

By analogy to MFRs, one can define an OFR resonance frequency width $w$ by
\begin{equation}
\label{eqn:Gamma_OFR}
   \frac{ \hbar \Gamma_\mathrm{OFR}(k)}{2k}  = a_\mathrm{bg} (-\hbar w ) \,.
\end{equation}
Note that $-a_\mathrm{bg} w$ is positive definite. In the limit $|\delta| \gg \gamma$ where we can ignore the decay of the resonance,  the scattering length due to an OFR is
\begin{equation}
\label{eqn:a_OFR_abg}
   a(\omega) = a_\mathrm{bg} - a_\mathrm{bg} \frac{w}{\omega-(\omega_n+s_n I)} \,.
\end{equation}

The standard way to express the  $L_\mathrm{r} E_\mathrm{r}$ product in the case of an OFR is to define $E_\mathrm{r}$ to be the known quantity $\hbar \gamma_\mathrm{m}$ and call the length parameter the ``optical length'' $l_\mathrm{opt}$~\cite{Ciuryo2005,Blatt2011},
\begin{equation}
\label{eqn:Gamma_OFR_lopt}
\frac{ \hbar \Gamma_\mathrm{OFR}(k)}{2k} = l_\mathrm{opt} (\hbar\gamma_{m}) \,.
 \end{equation}
\noindent For large detunings $|\delta| \gg \gamma$ the scattering length is
\begin{equation}
\label{eqn:a_OFR_lopt}
   a(\omega) = a_\mathrm{bg} + l_\mathrm{opt} \frac{\gamma_{m}}{\omega-(\omega_n+s_n I)} \,.
\end{equation}
A similar resonance length parameter has been defined for a decaying MFR by Hutson \cite{Hutson2007} by using the total decay width of the resonance for $E_\mathrm{r}$.

In the case of an OFR that decays only to the ground state, choosing $E_\mathrm{r}=\hbar \gamma_m$ has the advantage of eliminating the dipole strength from the expression for $ l_\mathrm{opt}$. Using $\hbar \gamma_a= 32\pi^3 d_a^2/3 \lambda_a^3$, the definition $\gamma_m = 2\gamma_a$, taking $V_\mathrm{opt}$ in Eq.~(\ref{eqn:Vopt}), and assuming the $R$-independent molecular dipole moments of Eqs.~(\ref{eqn:d0u}) or~(\ref{eqn:d1u}), we find
\begin{equation}
\label{eqn:l_opt}
  l_{\mathrm{opt}} =\frac{\Gamma(k)}{2 k \gamma_{m}} =  \frac{\lambda_{a}^{3}}{16 \pi c} \frac{| \langle n | E \rangle |^{2}}{k} I f_{\mathrm{rot}} \,,
\end{equation}
where $f_{\mathrm{rot}}$ is equal to 1 for $0_u$ states and 2 for $1_u$ states (due to the different rotational H{\"o}nl-London factors for parallel and perpendicular transitions).  Consequently, $L_r E_r$ is proportional to the product of a Franck-Condon factor and the square of the molecular electronic transition dipole moment.

Equation~(\ref{eqn:l_opt}) shows that $l_\mathrm{opt}$ varies linearly with PA laser power $I$.  The only molecular physics parameter it depends on is the free-bound Franck-Condon factor $| \langle n | E \rangle |^{2}$, which varies linearly with $k$ at small $k$. As an example, direct calculation of  $| \langle n | E \rangle |^{2}/k$ for the $J=0$ $n=-4$ $0_u$ level shows that this quantity decreases at a rate of 0.66\% per $\mu$K as $E/k_B$ ranges from 0 to 10 $\mu$K.  Thus, $l_\mathrm{opt}/I$ is only weakly dependent on collision energy in ultracold gases and may be approximated as a constant. Its weak variation with energy could be estimated from approximate theories based on the reflection approximation~\cite{Bohn1999} or the stationary phase approximation~\cite{Ciurylo2006}.

A useful way to compare the strengths of MFRs and OFRs is to use a dimensionless resonance ``pole strength'' parameter that applies to either case: $s_\mathrm{res} = L_\mathrm{r} E_\mathrm{r}/\bar{a}\bar{E}$, where $\bar{a}$ is the mean scattering length of the van der Waals potential~\cite{Gribakin1993} and $\bar{E}=\hbar^2/(2\mu \bar{a}^2)$ is the corresponding energy.  These are $\bar{a}=$ 71.8~$a_0$ and $\bar{E}/h=$ 7.97~MHz for $^{88}$Sr collisions.    Chin {\it et al.}~\cite{Chin2010} used $s_\mathrm{res}$ to characterize and classify MFRs according to whether $s_\mathrm{res} >1$ (open channel dominated) or $s_\mathrm{res} <1$ (closed channel dominated), where the former tends to be ``broad'' and the latter ``narrow.''  Thus we have
\begin{equation}
 s_\mathrm{res}^\mathrm{MFR} = \frac{a_\mathrm{bg}}{\bar{a}} \frac{ \Delta \delta \mu}{\bar{E}} \,, \,\,\,\,
  s_\mathrm{res}^\mathrm{OFR} = \frac{l_\mathrm{opt}}{\bar{a}} \frac{ \hbar \gamma_m}{\bar{E}} \,.
\end{equation}
One obvious difference between MFRs and OFRs is that the strength of an OFR can be controlled by increasing the PA laser power to increase $l_\mathrm{opt}$, whereas the strength of a MFR is fixed.  However, $l_\mathrm{opt}$  cannot be increased too much since the light scattering loss rate due to either atomic or molecular processes also increases with $I$~\cite{Blatt2011}.

Experimentally useful MFRs tend to have a pole strength parameter between unity and 100~\cite{Chin2010}. The width ratio $\hbar \gamma_m/\bar{E} =0.0019$ is much less than unity for the narrow OFRs near the intercombination line of $^{88}$Sr, so $s_\mathrm{res}^\mathrm{OFR}  \ll 1$ unless it can be compensated by making $l_\mathrm{opt}/\bar{a}$ very large compared to unity.  Thus, $^{88}$Sr OFRs tend to be weak, narrow, ``closed channel dominated'' resonances.  An interesting comparison is with the experimentally useful broad open channel dominated MFR of two $^{85}$Rb atoms at 155.2 G, for which $s_\mathrm{res}=$ 28~\cite{Chin2010}.  This is a decaying MFR in an excited spin channel~\cite{Kohler2006}, with a natural decay width of  $\gamma/(2\pi)=$ 5.0 kHz due to spin relaxation of the ``bare'' closed channel state of the resonance. The lifetime of 32 $\mu$s~\cite{Kohler2005} of this spin channel is comparable to that of the excited Sr $^3$P$_1$ state.  The major difference between the $^{85}$Rb MFR and $^{88}$Sr OFRs is the much smaller resonance strength $s_\mathrm{res}$ of the latter at intensities where the atomic light scattering is not harmful.

In contrast to $^{88}$Sr,  OFRs for the species $^{172} \!\!$~Yb  were found to have an $l_\mathrm{opt}$ on the order of $10^4 a_0$ at $I = 1 \mathrm{W/cm}^2$ for levels near 1~GHz atomic detuning \cite{Borkowski2009}. This implies that broad open-channel-dominated OFRs with $s_\mathrm{res} > 1$ may be realizable with $^{172} \!\!$~Yb.  It is not yet known whether OFRs might exist with $s_\mathrm{res} > 1$ for frequencies near the alkaline earth intercombination line in mixtures of alkaline earth species and alkali metal species.  This is a subject for future experimental and theoretical research.

\subsection{Multiresonance Theory}
\label{subsec:OFR_MFR_multires}

It is useful to compare isolated OFRs and MFRs since isolated resonance theory is widely utilized in both cases; however, both OFRs and MFRs exist as sets of resonances that interfere with one another, so it is instructive to compare multiresonance treatments of the two effects. Although a thorough treatment of a multiresonance OFR-MFR comparison could be the subject of an entire publication, in this section we provide an overview of such a comparison using results from multichannel quantum defect theory (MQDT).

When resonance interference is considered, significant qualitative differences between OFRs and MFRs emerge.  There are two sources of such differences. First, the molecular physics that determines the resonance strength is due to short range spin-dependent interactions for MFRs and long-range photoassociation for OFRs. Second, OFRs typically span many vibrational levels of the same closed channel molecular state whereas experimentally utilized MFRs are typically different spin components rather than a vibrational progression.

Sets of overlapping MFRs are well-studied for different alkali metal species~\cite{Naik2011,Berninger2013,Takekoshi2012,Tung2013,Repp2013,Jachymski2013}, and overlapping MFRs have recently been shown to be important for few-body physics~\cite{Wang2014}.  The interference of overlapping MFRs can be quantitatively explained by MQDT~\cite{Mies1984a,Mies1984b,Jachymski2013}, with which one can derive an $S$-matrix that is a multiresonance generalization of the isolated resonance formula in Eq. (\ref{eqn:bohn_s_matrix}).  We introduce the MQDT theory here to highlight the differences between overlapping OFRs and MFRs.

Considering one background channel and one closed channel, the background is characterized by the usual $E$-dependent phase shift $\eta_\mathrm{bg}(E)$~\cite{Jachymski2013,Mies1984a,Mies1984b,MQDTnote1}.  The closed channel c is characterized by a bound state phase function $\nu_\mathrm{c}(E-E_\mathrm{c})$. The energy $E_\mathrm{c}$ of the separated atoms in the closed channel is modified with the ``field tuning,'' which means varying the external magnetic field in the MFR case or the atomic detuning in the OFR case. Bound states of the closed channel exist where $\tan\nu_\mathrm{c}(E-E_\mathrm{c})=0$. Thus, the external field tuning moves the bound state spectrum relative to the background channel $E=0$ threshold, allowing bound states to be tuned across threshold.   The coupling between the background and the closed channel is characterized by the dimensionless MQDT parameter $s_\mathrm{res}$, which may also depend on the external fields.

If we follow Ref.~\cite{Jachymski2013} and express energies as $\epsilon=E/\bar{E}$ and $\kappa = k \bar{a}$, then the equivalent MQDT expression to Eq. (\ref{eqn:bohn_s_matrix}) for an MFR or an OFR can be written in a universal dimensionless form,
\begin{equation}
 S_{MQDT} =  S_\mathrm{bg}
  \left(1-i\frac{2\kappa s_\mathrm{res}}{(\epsilon_\mathrm{m}/\pi)\tan\nu_\mathrm{c}+\epsilon_\mathrm{shift}+i\epsilon_\mathrm{width}}\right)\, ,
\label{eqn:MQDT_S}
\end{equation}
where $S_\mathrm{bg}(\epsilon)=e^{2i\eta_\mathrm{bg}(\epsilon)}$ is the background term,
\begin{equation}
\epsilon_\mathrm{m}(\epsilon)=\frac{\pi}{\partial \nu_\mathrm{c}(\epsilon)/\partial \epsilon}
\end{equation}
represents the mean spacing between different eigenenergies, and  $\epsilon_\mathrm{shift}(\epsilon)$ and $\epsilon_\mathrm{width}(\epsilon)=\frac12 \hbar \gamma/\bar{E}+\kappa s_\mathrm{res}$ represent the respective shift and decay parts of the complex energy of the interacting, decaying resonance.  The shift term $\epsilon_\mathrm{shift}(\epsilon)$ will scale as $s_\mathrm{res}$, and both of these quantities are only very weakly dependent on energy near threshold.   Here we need to view the MQDT parameter $s_\mathrm{res}$ as a continuous function of the external field that defines the Hamiltonian.

Using Eq.~(\ref{eqn:MQDT_S}), we can now describe some key differences of OFRs and MFRs.  These come from the variation with field strength of both the numerator and the denominator of Eq.~(\ref{eqn:MQDT_S}).    In the MFR case, it is an excellent approximation to take $s_\mathrm{res}$ to be a constant, independent of $B$ and $E$, since the interactions that determine this parameter are short range, where $R \ll \bar{a}$ and the energy scale is large.  Consequently, the matrix element that sets the magnitude of $s_\mathrm{res}$ is independent of small field tuning.  On the other hand for an OFR, $s_\mathrm{res}$ scales linearly with laser power, and we must also think of $s_\mathrm{res}(\omega)$ as being highly sensitive to field tuning, since the optical coupling is determined by the Condon point at very long range (on the order of $\bar{a}$ or larger). The Condon point varies rapidly with PA laser frequency, so the crossing structure of the field-dependent Hamiltonian varies with laser frequency in a major way, changing the response of the system to the optical field.  Another way of thinking about this variation is that for an isolated resonance $s_\mathrm{res}$ is proportional to a Franck-Condon factor, which will vary rapidly from level to level in the closed channel, so that the general MQDT coupling parameter can not be taken as a field-tuning-independent parameter \cite{MQDTnote2}.

Secondly, note that the proper MQDT expression in the denominator of Eq.~(\ref{eqn:MQDT_S}) that contains the effect of field tuning is the $\tan\nu_\mathrm{c}$ term, which vanishes at resonance poles.  To get the normal isolated resonance approximation near a tunable eigenenergy $\epsilon_\mathrm{res}$, as in Eq.~(\ref{eqn:bohn_s_matrix}), it is necessary to expand this function in a Taylor series as~\cite{Mies1984b,Jachymski2013}
\begin{equation}
  \tan\nu_\mathrm{c}(\epsilon-\epsilon_\mathrm{c}) \approx \frac{\partial \nu_\mathrm{c}(\epsilon-\epsilon_\mathrm{c})}{\partial \epsilon}{|}_{\epsilon=\epsilon_\mathrm{res}} \left ( \epsilon - \epsilon_\mathrm{res} \right ) \,.
  \label{eqn:tan_nu}
\end{equation}
In the ultracold case, $\epsilon$ tends to remain very close to 0 but $\epsilon_{res}$ is varied by tuning the field, so the expansion in Eq.~(\ref{eqn:tan_nu}) should be made near $\epsilon=0$.  This linearizing approximation is normally quite good as long as the range of expansion $\epsilon-\epsilon_\mathrm{res}$ remains small compared to the mean spacing $\epsilon_\mathrm{m}$ to adjacent levels.  This is normally the case for MFRs, where the widths of even broad resonances tend to be small compared to the distance to the next vibrational level in the same spin channel \cite{MQDTnote3}. On the other hand, it is common to observe a series of vibrational levels of the same electronic state in the OFR case.  In this case, the  $\tan\nu_\mathrm{c}$ must be left unexpanded and $\nu_\mathrm{c}$ treated as a continuous function of field tuning if multiple resonances are present \cite{MQDTnote4}.

Reference~\cite{Jachymski2013} showed how to extend the MQDT formalism for Eq.~(\ref{eqn:MQDT_S}) to multiple spin channels with overlapping resonances.  Generally, for the reasons discussed above $s_\mathrm{res}$  for each separate channel can be well-approximated as a $E$- and $B$-independent constant, and the linearizing approximation in Eq.~(\ref{eqn:tan_nu}) is used for detunings spanning multiple spin channels.  Then the generalization of Eq.~(\ref{eqn:tan_nu}) gives a sum of resonance terms similar to that in the pole term of Eq.~(\ref{eqn:tan_nu}), where there is a global background scattering length $a_\mathrm{bg}$ for all channels and the shift terms in each denominator depend on all the poles simultaneously.  The formula can be transformed to a form where each narrow resonance can be viewed as an isolated resonance having a ``local'' (in field tuning) background modified from the global one by nearby interfering resonances.  An extension to OFRs from different electronic states may not be possible, because of the rapid variation of MQDT parameters with field tuning.  Furthermore, the MFR theory should not be used for different members of the same vibrational series because of the inability to linearize the $\tan\nu_\mathrm{c}$ function across two or more vibrational levels.  It may be possible to develop some approximations appropriate to the OFR case, but meanwhile the isolated resonance approximation or CC calculations remain the best tools for understanding OFRs.

\section{Elastic and Inelastic Collisions}
\subsection{OFR isolated resonance formulas}
\label{subsec:iso_res_formulas}

For detunings as large as hundreds of linewidths from the resonance center (but smaller than the separation between resonances), the complex scattering length $\alpha(k,\omega,I)$ and the elastic or inelastic collision rate coefficients derived from the $S$-matrix of Eq.~(\ref{eqn:bohn_s_matrix}) are in excellent agreement with the full CC calculations. In this regime, Eq.~(\ref{eqn:bohn_s_matrix}) fully describes an isolated OFR as a function of energy, detuning, and intensity.

The expression in Eq.~(\ref{eqn:bohn_s_matrix}) gives the isolated resonance approximation to the entrance channel loss probability $P_\textrm{loss}$ in terms of only two parameters, $l_\mathrm{opt}$ and the resonance position $E_\mathrm{res}$:

\begin{equation}
\label{eqn:loss}
  P_\textrm{loss} = 1 - |S(k)|^2 = \frac{2 k l_\textrm{eff}}{D^2 + \frac14(1+ 2 k l_\textrm{eff})^2} \,,
\end{equation}
where
\begin{eqnarray}
 l_\textrm{eff} &=& l_\mathrm{opt} (\gamma_m/\gamma), \label{eqn:x}  \\
 D &=& (E -E_\mathrm{res})/\hbar \gamma  \label{eqn:D} \,.
\end{eqnarray}

Here the effective optical length $l_\textrm{eff}$ determines the resonance coupling strength for the general case when $\gamma >\gamma_m$. $P_\textrm{loss}$ determines the inelastic cross section in Eq.~(\ref{eqn:in_cross_section}) and thus the loss rate coefficient $K_{in}$ in Eq.~(\ref{eqn:loss_rate}). Note that for $2 k l_\textrm{eff} \gg 1$, Eq. \ref{eqn:loss} describes power broadening.

\begin{figure}
  \centering
  \includegraphics[width=\linewidth]{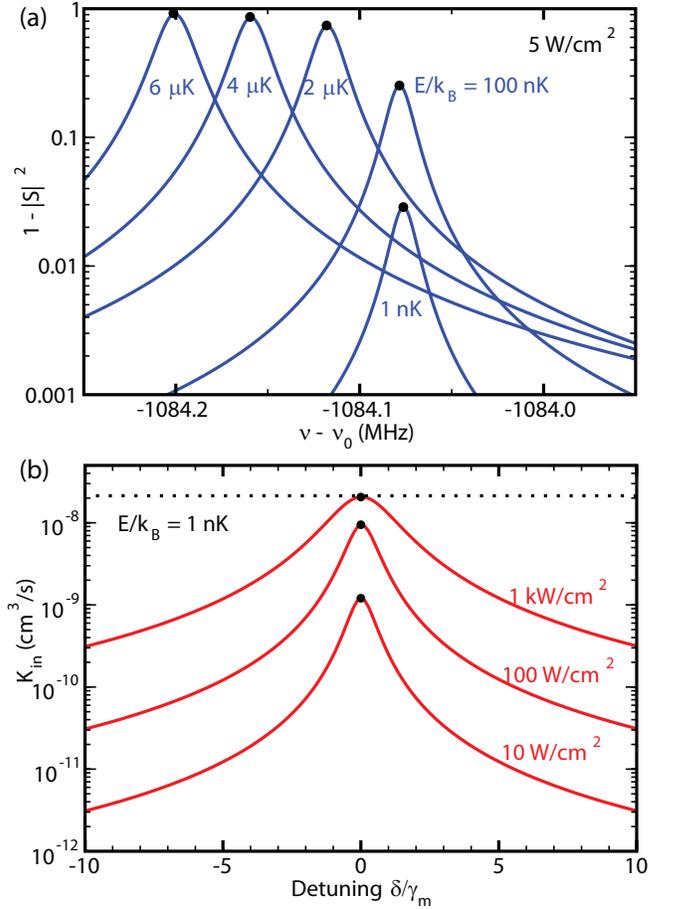}
    \caption{(Color online) (a) Coupled-channel calculated inelastic loss probability $1 - |S(k)|^2$ versus atomic detuning $\nu-\nu_0$ for different values of collision energy $E/k_B$. We have used $I=$ 5 W/cm$^2$ and the $0_u$ $J=1$ $n=-4$ feature, where $E_n/h = -1084.0763$ MHz. The black dots indicate the peak values calculated with $l_\mathrm{opt}=$161.5 a$_0$, $\gamma =\gamma_m$, and $\nu-\nu_0=(E_n-E)/h$ (where $D$ vanishes in Eq.~(\ref{eqn:loss})). Since $1 - |S(k)|^2$ calculated from the isolated resonance formula in Eq.~(\ref{eqn:loss}) is indistinguishable from the CC calculation on this graph, only the peak comparisons are shown by the black dots on the figure. Panel (b) Coupled channels calculated inelastic loss rate coefficient $K_{in}$ versus molecular detuning. Here we plot different values of PA laser intensity $I$. We use a collision energy $E/k_B=$ 1 nK and the $0_u$ $J=1$ $n=-4$ feature near $\nu - \nu_0 =$ $-1084$ MHz.  The black dots show the predictions of the analytic formula in Eq.~(\ref{eqn:loss}). The line wings beyond around $|\delta/\gamma_m| = 6$ scale linearly with power. The black dotted line represents the unitarity limit where $K_{in}$ saturates.}
  \label{fig:Iso_Ploss}
\end{figure}
Figures~\ref{fig:Iso_Ploss}(a) and ~\ref{fig:Iso_Ploss}(b) show the behavior of $P_\textrm{loss}$ as a function of detuning for different intensities and collision energies.  In Fig. \ref{fig:Iso_Ploss}(a), $P_\textrm{loss}$ is not saturated at low collision energy.  As collision energy increases, $P_\textrm{loss}$ broadens, and its peak value (when $E=E_\mathrm{res}$) approaches its upper bound of unity. This only occurs for red molecular detunings. Fig. \ref{fig:Iso_Ploss}(b) shows similar broadening in the inelastic rate coefficient $K_{in} \propto P_\mathrm{loss}$ as intensity is increased and collision energy is kept low. For large intensities, $K_{in}$ saturates at its value given by the unitarity limit. Note that according to Eq. \ref{eqn:Eres}, the intensity-dependent frequency shift of the resonance is accounted for in $\delta$. For temperatures in the $\mu$K range, thermal averaging of the line shape is essential to compare with experiment.  As is well-known~\cite{Jones2006}, PA lines have a pronounced red-blue asymmetry when $k_B T$ is larger than the natural width of the PA line.

The isolated resonance \textit{S} matrix can be used to derive the complex $k$-dependent scattering length $\alpha(k)$. Combining Eqs. (\ref{eqn:bohn_s_matrix}) and (\ref{eqn:scattering_length}),
\begin{equation}
   \alpha(k) = \alpha_\mathrm{bg}(k) + \frac{\frac{\hbar \Gamma(k)}{2k} \left ( 1 + k^2 \alpha_\mathrm{bg}(k)^2 \right )} {E-E_\mathrm{res} - k \alpha_\mathrm{bg}(k) \frac{\hbar \Gamma(k)}{2} + i \frac{\hbar \gamma}{2}} \,,
\end{equation}
where $\alpha_\mathrm{bg}(k)$ is found by using the background $S_\mathrm{bg}(k)=e^{2i\eta_\mathrm{bg}}$ in Eq.~(\ref{eqn:scattering_length}). Notice that this expression does not contain power broadening, which enters the elastic and inelastic cross sections through the $f(k)$ factor in Eqs.~(\ref{eqn:el_cross_section})-(\ref{eqn:f_factor}).

\begin{figure}
  \centering
  \includegraphics[width=\linewidth]{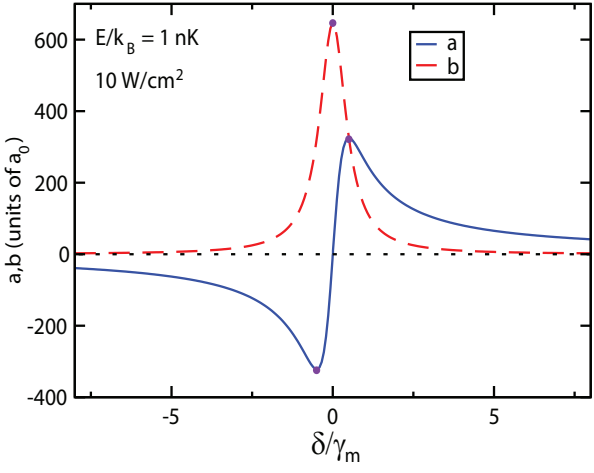}
  \caption{(Color online) Coupled channel calculation of $a$ and $b$ versus molecular detuning in line width units $\delta/\gamma_\mathrm{m}$, with $\gamma=\gamma_\mathrm{m}$. We use the $0_u$ $J=1$ $n=-4$ feature and 10W/cm$^2$ PA laser intensity.  The dots show the analytic predictions at the peak extrema using Eq.~(\ref{eqn:Iso_alpha}) with $l_\mathrm{opt}=323.2$ a$_0$.}
  \label{fig:Iso_a+b}
\end{figure}

In the limit that $k|\alpha_\mathrm{bg}| \ll 1$ (valid for a Bose-Einstein condensate), we obtain 

\begin{equation}
 \alpha = a_\mathrm{bg} + l_\mathrm{eff} \frac{\delta \gamma}{\delta^{2} + \gamma^{2}/4}  - \frac{i}{2}  l_\mathrm{eff} \frac{\gamma^{2}}{\delta^{2} + \gamma^{2} / 4} \,,
  \label{eqn:Iso_alpha}
\end{equation} 
where we have taken $\alpha_\mathrm{bg}=a_\mathrm{bg}$ to be real. Fig.~\ref{fig:Iso_a+b} illustrates the real and imaginary parts of $\alpha=a-ib$ calculated at $E/k_B=$ 1 nK.  The results from Eq.~(\ref{eqn:Iso_alpha}) are in excellent agreement with the CC calculations.  The peak values of $b=2 l_\mathrm{opt}$ at $\delta=0$ and of $a=\pm l_\mathrm{opt}$ at $\delta=\pm \gamma/2$ are also plotted in Fig.~\ref{fig:Iso_a+b}.

\subsection{Photoassociation}
\label{subsec:PA}

The isolated resonance approximation also describes photoassociation~\cite{Ciuryo2004}, the process by which two ground state atoms and a photon combine to form an electronically excited molecule~\cite{Jones2006}. This phenomenon can be used to measure $l_\mathrm{eff}$, which characterizes the strength of an OFR. This strength can be inferred from measurements of the trapped atom loss (into untrappable molecules) that results from driving a photoassociation resonance. Using Eqs. (\ref{eqn:in_cross_section}) and (\ref{eqn:loss_rate}), the inelastic rate coefficient, which describes molecule formation, is

\begin{equation}
\label{eqn:K_in}
K_{in} (\delta,l_\textrm{eff},k) = \frac{4 \pi \hbar}{\mu} \frac{\gamma^{2} l_\textrm{eff}}{(\delta + E/\hbar)^{2} + \frac{\gamma^2}{4} (1 + 2 k l_\textrm{eff})^{2}}.
\end{equation}

For a trapped ultracold thermal gas, one must introduce Boltzmann averages into the theory. To this end, we approximate that the PA laser interacts with an entire velocity class at each point in space within the trap. This approximation, which is good for large densities, means that one must momentum average $K_{in}$. Furthermore, we use the fact that photoassociation is a short-range phenomenon in the isolated resonance approximation~\cite{Bohn1999}; therefore, we do not perform a Boltzmann spatial average in this treatment.

The quantity $\overline{K}_{in}$, which is the momentum-averaged $K_{in}$, is given by

\begin{eqnarray}
\label{eqn:Kin_averaged}
\overline{K}_{in} &=& \frac{1}{\pi^3 p_0^6} \int \!\! d^3 \vec{p}_1 \!\! \int \!\! d^3 \vec{p}_2 \, e^{-(p_1^2 + p_2^2)/p_0^2} K_{in}(\delta,l_\textrm{eff},k) \notag \\
&=& \frac{4}{\sqrt{\pi}k_{th}^3} \int_0^\infty \!\! dk \, k^2 \, e^{-k^2/k_{th}^2} K_{in}(\delta,l_\textrm{eff},k),
\end{eqnarray}

\noindent where $p_0 = \sqrt{2 m k_B T}$, $k_{th} = \sqrt{2 \mu k_B T}/\hbar$, $\vec{p}_1$ and $\vec{p}_2$ are the momenta of the two collision partners, and $|\vec{p}_1 - \vec{p}_2| = \hbar k$.

In order to use $\overline{K}_{in}$ to describe trap loss due to photoassociation, we must understand what fraction of molecules is ejected from the trap. As mentioned in Section \ref{subsec:CC_formulation}, we approximate that every photoassociated molecule is lost to the trap. We have numerically studied this approximation, finding that it is good to 1\% for all molecular states except for the least-bound $0_u$ level. In this case, the evolution of the in-trap atomic density is

\begin{equation}
\label{eqn:two_body_rate_equation}
\dot{\rho} = -2 \overline{K}_{in} \frac{\rho^{2}}{2} - \frac{\rho}{\tau},
\end{equation}

\noindent where $\rho$ is the atomic density and $\tau$ is the one-body lifetime (due to loss mechanisms such as background gas molecules and atomic light scatter). Here the $\rho^{2}$ term arises from the number of pairs in an $N$-particle sample, $N(N-1)/2 \simeq N^{2}/2$. The signal in photoassociation experiments is typically the atom number $N$ after the application of the PA laser \cite{Blatt2011,Escobar2008}, given by integrating the solution to Eq. (\ref{eqn:two_body_rate_equation}) over space,

\begin{equation}
\label{eqn:number_signal}
N(\delta,l_\mathrm{eff}) = \int d^{3} \vec{r} \frac{\rho_{0} (\vec{r}) \,\, e^{-t_{PA} / \tau}}{1 + \overline{K}_{in}(\delta,l_\textrm{eff}) \,\, \rho_{0}(\vec{r}) \,\, \tau (1 - e^{-t_{PA}/\tau})}.
\end{equation}

\begin{figure}
  \centering
  \includegraphics[width=\linewidth]{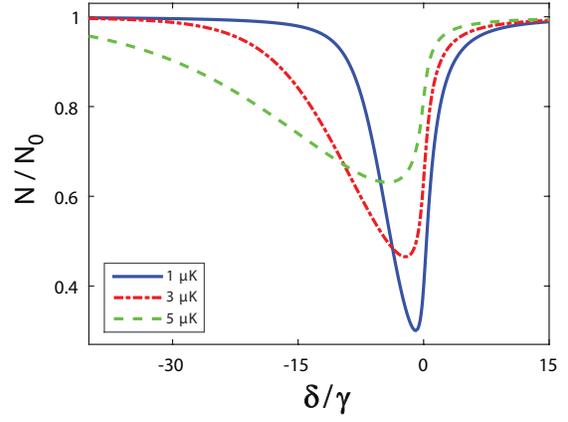}
  \caption{(Color online) Trap loss, given by Eq. \ref{eqn:number_signal}, as a function of detuning for various cloud temperatures. Here $\gamma = \gamma_\textrm{m}$, $l_\textrm{eff} = 100 a_0$, $t_{PA} = 200 \mathrm{ms}$, and we take the limit of $\tau \gg t_{PA}$. We also assume a spherical trapped cloud with $N_0 = 6 \times 10^4$ atoms and a 20 $\mu$m r.m.s. radius. Note the broadening toward red detunings, which is a result of momentum averaging $K_{in}$.}
  \label{fig:signal}
\end{figure}

\noindent Here $t_{PA}$ is the pulse duration of the PA laser and $\rho_{0} (\vec{r})$ is the in-trap density before the PA laser is applied. Fig. (\ref{fig:signal}) depicts $N(\delta,l_\mathrm{eff})$ for different temperatures and detunings. The density $\rho_{0}(\vec{r})$ can be determined by fitting experimental in-trap absorption images to a 3D Gaussian distribution (Ref. \cite{Blatt2011}, supplementary online material).

Unless a magic wavelength trap is employed, the ac Stark shift from optical traps causes a position-dependent frequency shift of the atomic resonance \cite{Blatt2011, Escobar2008}. For photoassociation experiments, this effect results in a broadening of the lineshape feature in the signal $N$. To model this broadening, one must understand both the atomic response to the optical trap and the intensity profile $I_{trap}(\vec{r})$ of the trap laser (Ref. \cite{Blatt2011}, supplementary online material). In the case of \Sr, the $^1S_0$ and $^3P_1$ polarizabilities are known well enough to calculate the differential polarizability, $\alpha_{^3 \! P_1}(\omega_{trap}) - \alpha_{^1 \! S_0}(\omega_{trap})$, to better than 10\% for typical trap laser wavelengths (such as 1064 nm).

\begin{figure}
  \centering
  \includegraphics[width=\linewidth]{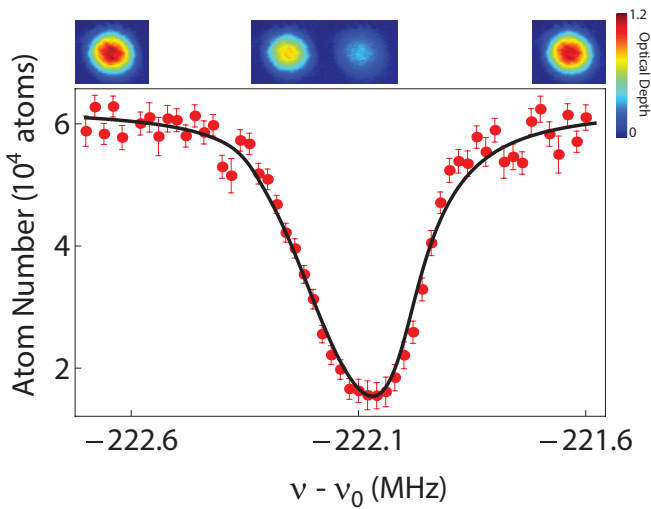}
  \caption{(Color online) Atom number data from Ref. \cite{Blatt2011} as a function of the PA laser detuning from the atomic resonance. The false color pictures above the plot are the measured optical depths corresponding to the data points directly below the centers of the pictures. These measurements were performed in a non-magic wavelength trap, which caused broadening toward blue detunings since $\omega_{Stark}(0) = 2 \pi \times 327 \,\, \mathrm{kHz}$. The solid line is a fit using Eq.~(\ref{eqn:number_signal}) with $\omega_{Stark}(\vec{r})$ included in $\overline{K}_{in}$. The quantities $T$ and $l_\mathrm{eff}$ were free parameters in the fit.}
  \label{fig:inelastic_loss_frequency}
\end{figure}

One can model $I_{trap}(\vec{r})$ with the Gaussian beam equation, using parametric heating measurements to obtain the beam waists \cite{Savard1997}. This broadening can be included in the photoassociation signal by adding a Stark shift,

\begin{equation}
\omega_{Stark}(\vec{r}) = -\frac{1}{2 \hbar \epsilon_0 c} \left[ \alpha_{^3 \! P_1}(\omega_{trap}) - \alpha_{^1 \! S_0}(\omega_{trap}) \right] I_{trap}(\vec{r}),
\end{equation}
to $\delta$ in Eq. \ref{eqn:K_in} and then carrying this term through to Eq. (\ref{eqn:number_signal}). With the trap ac Stark shift accounted for, we can fit experimental photoassociation data (Fig. \ref{fig:inelastic_loss_frequency}). For these fits it is necessary to approximate the integrals in Eqs. (\ref{eqn:number_signal}) and (\ref{eqn:Kin_averaged}) as sums (Ref. \cite{Blatt2011}, supplementary online material).

A full treatment of the photoassociation lineshape would include Doppler broadening. However, according to theoretical studies of narrow-line photoassociation~\cite{Ciuryo2004}, since a $T$ of a few $\mu$K (typical for narrow-line laser cooled \Sr{}) is greater than the PA laser photon recoil temperature, Doppler broadening is negligible compared to the momentum broadening shown in Fig. \ref{fig:signal}. We numerically checked whether Doppler effects are significant for our analysis, and we find that Doppler broadening can only be neglected in the fits of Fig. \ref{fig:inelastic_loss_frequency} due to the presence of both a large momentum broadening and an appreciable Stark shift broadening toward blue detunings.

Treating collisions in this manner breaks down when elastic processes become important. The elastic-to-inelastic collision ratio is approximately

\begin{equation}
\label{eqn:collision_ratio}
\overline{K_{el} (k) / K_{in} (k)} \simeq 2 \overline{k}
l_\textrm{eff} = 2 k_{th} l_\textrm{eff},
\end{equation}

\noindent where the overline denotes thermal averaging. In deriving this formula, we made the approximation $e^{2 i \eta_{bg}} \simeq 1$, which is acceptable for the above estimate since $a_{bg}$ is only -1.4 $a_{0}$. Therefore, when $l_\textrm{eff} \sim 1 / 2 k_{th}$, elastic collisions must be treated.

\subsection{Elastic Collisions}
\label{subsec:elastic_collisions}

If photoassociation can be minimized and elastic collision rate modifications can be made large, the OFR effect could be used to manipulate atomic interactions with relatively little particle loss. The usefulness of such manipulations is evident from experiments based on the MFR effect, which has proved to be a very fruitful technique that is central to numerous experiments \cite{Chin2010}. To access this regime in a quantum degenerate gas (for which $k \rightarrow 0$), the ratio of the optically modified elastic scattering length to the inelastic scattering length, $\left[ a(0)-a_{bg} \right]/b(0) = 2 \delta/\gamma$, must be much greater than unity. For a thermal gas, the rate coefficients determine the relevant limit, $\langle K_{el} (k) / K_{in} (k) \rangle \gg 1$, which implies via Eqn.\ref{eqn:collision_ratio} that $l_\textrm{eff} \gg 1/2 k_{th}$.

We estimate from the latter condition that a thermal \Sr{} gas at $T = 3~\mu\mathrm{K}$ (typical of narrow-line laser cooling) will require $l_\mathrm{eff}$ to be much greater than $400~a_0$. Using the isolated resonance approximation, large changes in elastic scattering were predicted to arise from an OFR based on the $n = -1$ vibrational state~\cite{Ciuryo2005}. This prediction required high PA laser intensity and very large molecular detunings from the $n = -1$ state. However, as our CC theory has shown, these conditions will not yield an effect comparable to a lossless MFR because the requisite detunings are larger than the separation between resonances. Instead, the OFR physics is determined by the nearest resonance to a given detuning (Section \ref{subsec:CC_Results}).

Experimentally, elastic collisions can be studied using cross-dimensional thermalization. For instance, in Ref. \cite{Monroe1993}, a trapped atomic gas was prepared in a nonequilibrium state using parametric heating in 1D of the trap. Due to elastic collisions, the authors observed thermalization of the non-equilibrium gas. The thermalization time of the gas can be calculated from a simple treatment based on Enskog's equation or a full molecular dynamics simulation~\cite{Goldwin2005}. These treatments show that, on average, each particle participates in about three elastic collisions events during the $1/e$ thermalization time.

Elastic collisions induced by OFRs were experimentally studied in Ref.~\cite{Blatt2011} using cross-dimensional thermalization. In this work, an OFR was accessed in a trapped \Sr{} gas prepared in a non-thermal state. In the absence of OFR-induced collisions, this gas would not thermalize over experimental time scales because of the negligible $a_\mathrm{bg} = - 1.4 a_0$ in \Sr{}. With a PA laser applied, clear temperature changes were observed as a function of $\nu - \nu_0$. To understand whether these observations were caused by elastic collisions, we apply our theory to the experimental data.

\begin{figure}
  \centering
  \includegraphics[width=\linewidth]{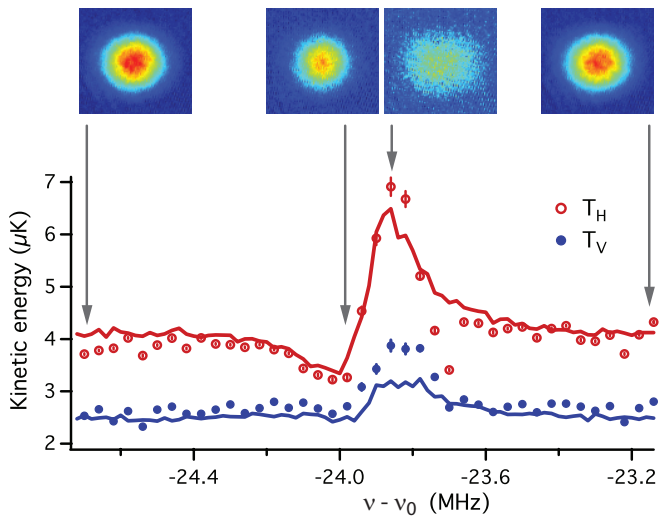}
  \caption{(Color online) Temperature data from Ref.~\cite{Blatt2011} for horizontal (H) and vertical (V) trap eigenaxes. The solid lines are the results of a Monte Carlo simulation including elastic and inelastic collisions.}
  \label{fig:simulation}
\end{figure}

Since the atomic samples in this measurement never reached thermal equilibrium over the time scale of the experiment, it is not possible to study the resulting data analytically. Instead, a time-dependent simulation of the phase-space density is necessary to understand the data quantitatively. Our analysis uses a numerically efficient method due to Bird~\cite{Bird94} for simulating collisions between thousands of particles. The method discretizes the trap volume into small ``collision volumes'' containing much less than one particle on average. If there is more than one particle in a collision volume, the probabilities $P_{el}$ and $P_{in}$ of elastic and inelastic collisions between these atoms is calculated as $P_{el/in} = |\vec{v}_1 - \vec{v}_2| \sigma_{el/in} t_{step}/V$. Here $\vec{v}_1$ and $\vec{v}_2$ are the velocities of the two atoms, $t_{step}$ is the time step of the simulation, $V$ is the collision volume, and the cross sections are given by

\begin{eqnarray}
\label{eqn:collision_cross_sections}
\sigma_\mathrm{el} &=& \frac{8 \pi}{1 + k^2 a_\mathrm{bg}^2}
\frac{\left[l_\mathrm{eff} \gamma + a_\mathrm{bg} (\delta +
\frac{E}{\hbar})\right]^2 + a_\mathrm{bg}^2
\gamma^2/4}{(\delta + E/\hbar)^2 + \frac{\gamma^2}{4}
(1 + 2 k l_\mathrm{eff})^2} \\
\sigma_\mathrm{in} &=& \frac{4 \pi l_\mathrm{eff}}{k} \frac{\gamma^2}{(\delta
+ E/\hbar)^2 + \frac{\gamma^2}{4} (1 + 2 k
l_\mathrm{eff})^2}.
\end{eqnarray}

If an inelastic collision occurs, both particles are removed from the simulation. For an elastic collision event, the particles' velocity vectors are rotated using a random rotation matrix. Between each of these collision steps, the particles evolve in the trap potential using an embedded Runge-Kutta method. We have checked this simulation against known results for cross-dimensional thermalization in harmonic traps and have also confirmed that, with elastic collisions removed, the simulation reproduces the results of photoassociation theory of the previous section~\cite{Blatt2011thesis}.

Fig. \ref{fig:simulation} depicts temperature data from the experiments of Ref.~\cite{Blatt2011} as well as our simulation results. Our simulation tells us that the temperature peaks for certain detunings because the PA laser is driving photoassociative loss of the coldest atoms, resulting in antievaporative heating. We also find that without including elastic collisions in our simulation, the simulation does not predict the dip in the horizontal temperature apparent in the data. Therefore, we conclude that this temperature dip indicates partial thermalization of the gas. The fact that antievaporation and thermalization have different detuning dependence arises because the elastic and inelastic collision rates average differently over the collision momentum $k$.

The interplay between elastic and inelastic processes is sensitive to the value of $\gamma$ used in the simulation. The simulation only agrees with the experimental data if we set $\gamma = 2\pi\times 40~\mathrm{kHz} = 2.7 \gamma_m$. This leads us to conclude that the OFR effect in \Sr{} is broadened beyond the natural spontaneous decay of the ${^{88}\mathrm{Sr}_2}$ molecules. Extra broadening has also been seen in other \Sr{}~\cite{Zelevinsky2006, Yan2013a} and Rb~\cite{Theis2004} OFR experiments.

\section{Summary and Conclusion}

We have developed CC and isolated resonance theories of OFRs. The CC theory predicts resonance interference for detunings between OFRs, causing the OFR effect to vanish between resonances. We conclude that OFR experiments have a ``nearest resonance'' constraint, meaning that the OFR effect is dictated by the nearest resonance to a given detuning. The isolated resonance theory agrees with the more complete CC theory for detunings near a molecular resonance. In this regime, it is possible to use the simpler isolated resonance theory to model photoassociation and OFR measurements and fit the data from these experiments. Such models have shown a broadening beyond the expected linewidth of the molecular state accessed by an OFR.

\bibliography{references}

\begin{thebibliography}{97}
\bibitem[{foo({\natexlab{a}})}]{footnote1}
\bibinfo{note}{The actual values of $C_{6,0u}$, $C_{6,1u}$ and $R_{0,0u}$,
  $R_{0,1u}$ are 3512.8 $E_h a_0^6$ and 3774 $E_h a_0^6$ and 6.867588 a$_0$ and
  6.525376 a$_0$ respectively, and $C_{3,0u}=2C_{3,1u} =2d_A^2=$ 0.015074 $E_h
  a_0^3$.}
\bibitem[{foo({\natexlab{b}})}]{footnote2}
\bibinfo{note}{The actual values of $C_{6,g}$, $C_{8,g}$ and $R_{0,g}$ are 3103
  $E_h a_0^6$, 379200 $E_h a_0^8$, and 9.39928 a$_0$ respectively.}
\bibitem[{MQD({\natexlab{a}})}]{MQDTnote1}
\bibinfo{note}{There are also two auxiliary threshold functions $C(E)$ and
  $\tan\lambda(E)$ defined by MQDT that need not concern us here.}
\bibitem[{MQD({\natexlab{d}})}]{MQDTnote4}
\bibinfo{note}{This is taken care of in our CC calculations, which do not make
  MQDT approximations but calculates the response of the system to the
  Hamiltonian for any given field tuning.}
\end{thebibliography}

\end{document}